\documentclass[english,biblatex]{lni}

\usepackage{graphicx}
\usepackage{subcaption}
\usepackage{tikz}

\usepackage{geometry}
\usepackage{booktabs}
\usepackage{tabulary}

\newcommand{\bra}[1]{\langle#1\rvert}
\newcommand{\ket}[1]{\lvert#1\rangle}
\newcommand{\Diag}{\mathrm{Diag}}

\begin{document}

\title{Connecting the Hamiltonian structure to the QAOA performance and energy landscape}

\author[Daniel Müssig \and Markus Wappler \and Steve Lenk \and Jörg Lässig]{Daniel Müssig\footnotemark[1] \and Markus Wappler\footnotemark[1] \and Steve Lenk\footnotemark[1]\textsuperscript{,}\footnotemark[2] \and Jörg Lässig\footnotemark[3]
\footnotetext[1]{Fraunhofer IOSB-AST, Dep. Cognitive Energy Systems, Görlitz, Germany \email{daniel.muessig@iosb-ast.fraunhofer.de}, \email{markus.wappler@iosb-ast.fraunhofer.de}}
\footnotetext[2]{German Aerospace Center (DLR), Institute of Quantum Technologies, Ulm, Germany \email{steve.lenk@dlr.de}}
\footnotetext[3]{University of Applied Sciences Zittau/Görlitz, Brückenstraße 1, 02826 Görlitz, Germany \email{jlaessig@hszg.de}}
}

\startpage{1} 
\editor{Herausgeber et al.} 
\booktitle{Name-der-Konferenz} 

\maketitle

\begin{abstract}
Quantum computing holds promise for outperforming classical computing in specialized applications such as optimization. With current Noisy Intermediate Scale Quantum (NISQ) devices, only variational quantum algorithms like the Quantum Alternating Operator Ansatz (QAOA) can be practically run. QAOA is effective for solving Quadratic Unconstrained Binary Optimization (QUBO) problems by approximating Quantum Annealing via Trotterization. Successful implementation on NISQ devices requires shallow circuits, influenced by the number of variables and the sparsity of the augmented interaction matrix. This paper investigates the necessary sparsity levels for augmented interaction matrices to ensure solvability with QAOA. By analyzing the Max-Cut problem with varying sparsity, we provide insights into how the Hamiltonian density affects the QAOA performance. Our findings highlight that, while denser matrices complicate the energy landscape, the performance of QAOA remains largely unaffected by sparsity variations. This study emphasizes the algorithm's robustness and potential for optimization tasks on near-term quantum devices, suggesting avenues for future research in enhancing QAOA for practical applications.
\end{abstract}
\begin{keywords}
Quantum Computing \and QAOA \and Max-Cut \and Sparsity \and Ising Hamiltonian 
\end{keywords}

\section{Introduction}
Quantum computing has the potential to significantly outperform classical computing in various specialized applications, one of which is combinatorial optimization. This use case involves identifying optimal solutions of an objective function over a finite domain. Numerous combinatorial optimization problems can be modeled as {\em Quadratic Unconstrained Binary Optimization (QUBO) problem} (cf. \cite{glover_tutorial_2019}). This is
\begin{equation}
\min x^TQx\,,\text{ subject to }x\in\{0,1\}^n\,,
\end{equation}
where $Q\in\mathbb{Z}^{n\times n}$. In the following, we assume, w.~l.~o.~g., $Q$ to be symmetric. QUBOs are well-suited for quantum computing, because they can subsequently be transferred 
into an {\em Ising Hamiltonian} and ultimately be solved by determining the minimum eigenvalue.

The Ising Hamiltonian is a model in physics used to describe interactions between spins on a lattice, including their interactions and response to external magnetic fields. The \textit{augmented interaction matrix} --- as will be defined in \cref{eq:augmented interaction matrix} --- combines both the interaction strengths between spins and the local magnetic fields into a single matrix, providing a comprehensive representation of the system.
A method of finding the minimum eigenvalue of an Ising Hamiltonian is the usage of a hybrid algorithm such as the {\em Variational Quantum Eigensolver (VQE)} or the {\em Quantum Alternating Operator Ansatz (QAOA)}, as motivated by \cite{hadfield_quantum_2019}. 
VQE is well-suited for solving optimization problems involving quantum variables, whereas QAOA is more appropriate for classical optimization problems (cf. \cite{moll_quantum_2018}). 

Hybrid algorithms leverage the capabilities of current-generation quantum computers by employing a shallow, parameterized quantum circuit run on a quantum machine, with parameter optimization performed on a classical computer. This iterative process continues until a sufficiently good parameter set is found.
To determine the ground energy of the Problem Hamiltonian $H_P$, using QAOA, a mixing Hamiltonian $H_M$ is required. A typical choice is $H_M = \sum_{j=1}^n X_j$, since its ground state, an equal superposition of all states, is easily prepared. Here and in the following, we denote by $U_j$ the unitary operator
$U_j=I^{\otimes(j-1)}\otimes U\otimes I^{\otimes(n-j)}$.
QAOA uses alternating applications of parameterized time evolution operators regarding $H_P$ and $H_M$. Thus, being
the {\em phase differentiator}
\begin{equation}
U_P(\gamma)=e^{-i\gamma H_P}\,,\;0\le\gamma<2\pi\,,
\end{equation}
and the {\em mixer}
\begin{equation}
U_M(\beta)=e^{-i\beta H_M}=\prod_{j=1}^n e^{-i\beta X_j}\,,\;0\le\beta<\pi\,.
\end{equation}
Consequently, a $p$-level QAOA circuit is initialized with the uniform superposition state and executes 
alternating applications of $U_P$ and $U_M$ for $p$ iterations to approximate adiabatic quantum annealing 
via Trotterization (cf. \cite{zhou_quantum_2020}):
\begin{equation}
U_M(\beta_p)U_P(\gamma_p)\cdot\ldots\cdot U_M(\beta_1)U_P(\gamma_1)H^{\otimes n}\ket{0}\,.
\end{equation}
Throughout this paper we limit ourselves to QAOA with $p=1$. 
The Problem Hamiltonian is derived from the objective function $f(x)=x^TQx$.
Firstly, we substitute $s_j=1-2x_j\,,\;1\le j\le n\,,$ to translate the objective
into
\begin{equation}\label{eq:x to s}
\hat{f}:\{+1,-1\}^n\rightarrow\mathbb{Z}\,,\quad
\hat{f}(s)=\sum\limits_{j=1}^nh_js_j+\sum\limits_{j<k}J_{jk}s_js_k\,.
\end{equation}
The Pauli-$Z$-gate has eigenvalues $\{+1,-1\}$, therefore ensuring that the
Problem Hamiltonian defined by
\begin{equation}\label{eq:s to Z}
H_P=\sum\limits_{j=1}^nC_j+\sum\limits_{1\le j<k\le n}C_{jk}\,,\quad
C_j=h_jZ_j\,,\;C_{jk}=J_{jk}Z_jZ_k\,,
\end{equation}
fulfills $H_P\ket{x}=f(x)$. QAOA will compute $\ket{\beta,\gamma}=U_M(\beta)U_P(\gamma)H^{\otimes n}\ket{0}$ and use sampling
to approximate the expectation value (ground energy) 
$\bra{\beta,\gamma}H_P\ket{\beta,\gamma}$. The classical part will use methods of numerical (continuous) optimization to 
optimize the expectation value over $\beta$ and $\gamma$.
\section{Related Work} 
\label{sec:related}
Ozaeta et al. \cite{ozaeta_expectation_2022} has given closed-form expressions
for calculating $C_j$ and $C_{jk}$. We make use of the so derived analytical
objective function to analyze the landscape of ground energies. Nevertheless we are still not able to analytically calculate optima, since we cannot easily calculate the derivative. Exemplary \cref{eq:analytic_cost_function} shows how to derive the analytical objective function for Max-Cut. Through the analytic objective function Ozaeta et al. have found that if we have coefficients greater than $1$, then the periodicity in $\gamma$ increases relative to the coefficient. Further, they found that if the coefficients of $h$ are greater than the ones of $J$ we push the minima and maxima outside and the peaks get narrower towards the middle. However, if the coefficients of $J$ are greater than those of $h$ then the shapes are changing towards the middle. Regarding the sparsity Ozaeta et al. have found that an increased number of edges per vertex leads to a narrowing of the peaks in $\gamma$ \cite{ozaeta_expectation_2022}. However, we will see later in the paper, that we can derive more information from the sparsity and structure of the Problem Hamiltonian.

In order to get more insight about an energy landscape without just looking at it is the usage of the Fourier transform. This method isolates the sinusoidal components \cite{stechly_connecting_2023}. The differences in the eigenvalues of the Problem Hamiltonian $H_P$ will be directly dependent on the frequencies in $\gamma$ direction. Using the Fourier transform St{\k e}ch{\l}y et al. \cite{stechly_connecting_2023} have introduced roughness metrics: Total Variation and Fourier density.
Total Variation is utilized in neural network loss landscapes to measure the extent of variation in a function across a finite domain by employing the first-order gradient. However, since the first derivative can not be calculated even with a complete analytical cost function, we have to use the numerical gradient approximation. A steeper slope shows us that the function oscillates more in this region. We use 1D slices of the energy landscape to generalize the Total Variation to multivariate functions. 
The Fourier density was renamed from Fourier sparsity by the authors. This metric is usually defined as the 0-norm. However, this implementation may lead to artificially large values, because of the finite precision of the floating point operations. Thus the authors have referred to the numerical sparsity utilizing the 1- and 2-norm.

\section{Max-Cut problem instances with different sparsities}
\label{sec:max-cut}
To investigate the influence of the sparsity of the augmented interaction matrix, we have to create problems with different sparsities. Max-Cut fits perfectly for this, since the number of edges in the graph determines the Hamiltonian sparsity. This NP-hard problem is fundamental in computer science and combinatorial optimization. Max-Cut has applications in various fields, including network design, statistical physics, and machine learning. To limit the number of graphs we use, but still have all possible sparsities, we use the non-isomorphic graph dataset \cite{mckay_non-isomorphic_nodate}.

For a given graph $G=(V=\{1,\ldots,n\}\,,\,E)$ a cut $V=V_0\dot\cup V_1$ can be represented by means of a
characteristic vector
\begin{equation}
x\in\{0,1\}^n\,,\;x_j=\begin{cases}
0\,,\;j\in V_0\,, \\
1\,,\;j\in V_1\,.
\end{cases}
\end{equation}
The (unweighted) Max-Cut problem can be defined as
\begin{equation}\label{eq:max-cut}
\max\sum\limits_{jk\in E}(x_j-x_k)^2\,,\text{ subject to }x\in\{0,1\}^n\,,
\end{equation}
confer \cite{glover_tutorial_2019}. This is a QUBO problem. Applying the same substitutions as described in
\cref{eq:x to s} and \cref{eq:s to Z}, we replace $x_j$ by $\displaystyle\frac{I-Z_j}{2}$ to obtain the Problem Hamiltonian.
To efficiently compute the energy landscapes below, we have made use of closed formulas for the expectation
value $\bra{\beta,\gamma} H_P\ket{\beta,\gamma}$ given by \cite{ozaeta_expectation_2022}. From \cref{eq:s to Z} we obtain
\begin{equation}
\bra{\beta,\gamma} H_P\ket{\beta,\gamma}=\sum\limits_{j=1}^n\bra{\beta,\gamma} C_j\ket{\beta,\gamma}+\sum\limits_{1\le j<k\le n}\bra{\beta,\gamma} C_{jk}\ket{\beta,\gamma}\,,
\end{equation}
and aforementioned closed formulas are
\begin{equation}
\begin{split}
    \bra{\beta,\gamma} C_j \ket{\beta,\gamma} &= 0\,,\\
    \bra{\beta,\gamma} C_{jk} \ket{\beta,\gamma} &= \frac{\sin(4\beta)}{2}\sin(2\gamma)[(\cos(2\gamma))^{d_j-1}+(\cos(2\gamma))^{d_k-1}]\\
    &-\frac{1}{2}(\sin(2\beta))^2(\cos(2\gamma))^{d_j+d_k-2f_{jk}-2}(1-(\cos(4\gamma))^{f_{jk}})\,.
\end{split}
\label{eq:analytic_cost_function}
\end{equation}
Here, $d_j$ and $d_k$ are the degrees of vertices $j$ and $k$, and $f_{jk}$ is the number of shared edges between vertices $j$ and $k$.
In this paper, we inspect the graphs given in \cref{tab:experiments} and shown in \cref{fig:graphs,fig:landscapes} of the non-isomorphic graph dataset. For better comparability and better clarity, we only use graphs with five vertices. Further, in our previous studies \cite{federer_application-oriented_2022,mussig_solving_2023} we started to see problems at four qubits. 

Using the definitions of $h_j$ and $J_{jk}$ from \cref{eq:x to s}, note that for Max-Cut $h_j=0,\,1\le j\le n$.
Instead we define the {\em augmented interaction matrix} $(m_{jk})_{j,k=1}^n$ with help of the main diagonal of the
Problem Hamiltonian (that is the objective values) and set $J_{jk}=J_{kj}$ for $j>k$:
\begin{equation}\label{eq:augmented interaction matrix}
m_{jk}=\begin{cases}
J_{jk}\,,\;j\neq k\,,\\
(H_P)_{jj}\,,\;j=k\,.\\
\end{cases}
\end{equation}
The reason for choosing the specific graphs is reflected in their special structures shown in \cref{sec:analytical}. In order to analyze these selected graphs effectively, it is crucial to understand the concept of matrix sparsity, which is a key characteristic of the matrices derived from these graphs.
\begin{table}
    \hspace*{-0.2cm}
    \centering
    \begin{tabulary}{\linewidth}{@{}JJCC>{\hsize=0.7cm}C@{}}
        \toprule
        Exp. & Analytic Cost Function & Total Variation & Fourier Density & Sparsity\\
        \midrule
        1 & {\scriptsize $0.5*\sin(4.0*\beta)*\sin(1.0*\gamma) - 0.5$} & $3.64$ & $3.98$ & $92\%$\\
        2 & {\scriptsize $0.5*(\cos(1.0*\gamma) + 1)*\sin(4.0*\beta)*\sin(1.0*\gamma) - 1.0$} & $2.58$ & $3.34$ & $84\%$\\
        3 & {\scriptsize $0.75*(\cos(1.0*\gamma)**2 + 1)*\sin(4.0*\beta)*\sin(1.0*\gamma) - 1.5$} & $6.60$ & $3.31$ & $76\%$\\
        7 & {\scriptsize $0.75*(1 - \cos(2.0*\gamma))*\sin(2.0*\beta)**2 + 0.375*\cos(4.0*\beta - 2.0*\gamma) - 0.375*\cos(4.0*\beta + 2.0*\gamma) - 1.5$} & $8.43$ & $5.32$ & $76\%$\\
        13 & {\scriptsize $1.0*\sin(4.0*\beta)*\sin(2.0*\gamma) - 2.0$} & $7.78$ & $2.54$ & $68\%$\\
        18 & {\scriptsize $-0.3125*\cos(4.0*\beta) - 0.09375*\cos(2.0*\gamma) - 0.1875*\cos(4.0*\gamma) - 0.03125*\cos(6.0*\gamma) + 0.015625*\cos(4.0*\beta - 6.0*\gamma) + 0.21875*\cos(4.0*\beta - 4.0*\gamma) + 0.671875*\cos(4.0*\beta - 2.0*\gamma) - 0.578125*\cos(4.0*\beta + 2.0*\gamma) - 0.03125*\cos(4.0*\beta + 4.0*\gamma) + 0.015625*\cos(4.0*\beta + 6.0*\gamma) - 3.1875$ } & $11.81$ & $3.14$ & $44\%$\\
        23 & {\scriptsize $1.5*\sin(2.0*\beta)**2*\sin(2.0*\gamma)**2 + 3.0*\sin(4.0*\beta)*\sin(1.0*\gamma)*\cos(1.0*\gamma)**2 - 3.0$} & $11.97$ & $4.56$ & $52\%$\\
        33 & {\scriptsize $2.5*(1 - \cos(2.0*\gamma)**3)*\sin(2.0*\beta)**2 + 5.0*\sin(4.0*\beta)*\sin(1.0*\gamma)*\cos(1.0*\gamma)**3 - 5.0$} & $19.80$ & $5.42$ & $20\%$\\
        \bottomrule
    \end{tabulary}
    \caption{Overview over the chosen graphs, their analytic cost function, total variation, fourier density, and sparsity.}
    \label{tab:experiments}
\end{table}
Matrix sparsity refers to the proportion of zero elements in a matrix compared to its total number of elements. A sparse matrix contains a high percentage of zero elements, while a dense matrix has few or no zero elements. The formula for computing the sparsity is given by 
\begin{equation}
    \text{sparsity}\Big((m_{jk})_{j,k=1}^n\Big) = \frac{\#\{m_{jk}=0:1\le j,k\le n\}}{n^2}\,.
    \label{eq:sparsity}
\end{equation}

\section{Analysis}
%
%
\label{sec:analytical}
We begin our analysis of energy landscapes by studying their symmetries. In fact, symmetries could be directly derived from \cref{eq:analytic_cost_function}.
In order to explain symmetries by problem properties, we will instead use the mixing operator
$U_M$ and the phase differentiation operator $U_P$.
The QAOA quantum circuit with one layer ($p=1$) reads
\begin{equation}
\ket{\beta,\gamma}
=\prod\limits_{j=1}^n e^{-i\beta X_j}\cdot U_P(\gamma)H^{\otimes n}\ket{0}\,.
\end{equation}
Considering a Max-Cut solution. Swapping the partition set for all nodes flips all bits
of the characteristic vector x, but does not change the objective value.
Hence,
\begin{equation}
\begin{split}
\ket{\beta,\gamma}
&=\prod\limits_{j=1}^n e^{-i\beta X_j}\cdot X^{\otimes n}U_P(\gamma)H^{\otimes n}\ket{0}\\
&=\prod\limits_{j=1}^n e^{-i(\beta+\pi/2) X_j}\cdot U_P(\gamma)H^{\otimes n}\ket{0}\,.
\end{split}
\end{equation}
This shows that the parameter $\beta$ is $\frac{\pi}{2}$-periodic and explains the
translational symmetry in $\beta$-direction of all example figures. 

Symmetries along $\gamma$ can be obtained by examining the phase differentiator
$U_P(\gamma)=\Diag(e^{-i\gamma f(0)},\ldots,e^{-i\gamma f(2^n-1)})$ with
$f(x)$ being the value of the respective cut. 
Let $\delta=\gcd(f(0),\ldots,f(2^n-1))$ be the greatest common divisor of all cut values.
Because of
\begin{equation}
e^{-i\gamma f(x)}=e^{-i(\gamma+2\pi/\delta)f(x)}\,,\;0\le x\le 2^n-1\,,
\end{equation}
the parameter $\gamma$ is $\frac{2\pi}{\delta}$-periodic in this case.
Evaluating examples Exp 13, Exp 7, Exp 18, and Exp 33, we obtain that all cuts
contain an even number of edges. Therefore, we have $\pi$-periodicity of $\gamma$, and
the respective figures show accordingly translational symmetry in $\gamma$-direction. 

Now, we analytically inspect the influences of different sparsities and structures within the augmented interaction matrix. 
In simpler graphs such as Exp 1 (\cref{fig:exp1}) and Exp 7 (\cref{fig:exp7}), with fewer edges and less connectivity, the energy landscapes show distinct and somewhat isolated wells, reflecting limited interactions between specific clusters of nodes. For example, Exp 1, with its disjoint sets of nodes, produces an energy landscape with distinct, separate energy wells. Similarly, Exp 7, despite having a few more edges, still shows somewhat isolated energy wells, indicating that the nodes are not as extensively interacting. These simpler structures result in energy distributions where the influence of individual edges is more apparent, with less overlap and interaction between different energy minima.
\begin{figure}
    \centering
    \begin{subfigure}[t]{0.24\textwidth}
        \raisebox{-\height}{\begin{tikzpicture}[node distance={15mm}, thick, main/.style = {draw, circle}] 
\node[main, fill=lightgray] (0) {$0$}; 
\node[main, fill=lightgray] (1) [above right of=0] {$1$}; 
\node[main, fill=lightgray] (2) [right of=1] {$2$}; 
\node[main, fill=lightgray] (3) [below right of=0] {$3$}; 
\node[main, fill=lightgray] (4) [right of=3] {$4$};
\draw (0) -- (4); 
\end{tikzpicture}}
        \caption{Exp 1 Graph}
    \end{subfigure}
    \hfill
    \begin{subfigure}[t]{0.24\textwidth}
        \raisebox{-\height}{\begin{tikzpicture}[node distance={15mm}, thick, main/.style = {draw, circle}] 
\node[main] (3) {$3$}; 
\node[main] (0) [above right of=3] {$0$}; 
\node[main] (4) [below right of=3] {$4$};
\node[main] (1) [right of=0] {$1$}; 
\node[main] (2) [right of=4] {$2$}; 
\draw (0) -- (3); 
\draw (0) -- (4); 
\draw (3) -- (4); 
\end{tikzpicture} }
        \caption{Exp 7 Graph}
    \end{subfigure}
    \hfill
    \begin{subfigure}[t]{0.24\textwidth}
        \raisebox{-\height}{\begin{tikzpicture}[node distance={15mm}, thick, main/.style = {draw, circle}] 
\node[main, fill=lightgray] (4) {$4$};
\node[main, fill=lightgray] (0) [above right of=4] {$0$}; 
\node[main, fill=lightgray] (1) [below right of=4] {$1$}; 
\node[main, fill=lightgray] (2) [right of=0] {$2$}; 
\node[main, fill=lightgray] (3) [right of=1] {$3$}; 

\draw (0) -- (4); 
\draw (1) -- (4); 
\end{tikzpicture}}
        \caption{Exp 2 Graph}
    \end{subfigure}
    \hfill
    \begin{subfigure}[t]{0.24\textwidth}
        \raisebox{-\height}{\begin{tikzpicture}[node distance={15mm}, thick, main/.style = {draw, circle}] 
\node[main] (0) {$0$}; 
\node[main] (1) [above right of=0] {$1$}; 
\node[main] (2) [right of=1] {$2$}; 
\node[main] (3) [below right of=0] {$3$}; 
\node[main] (4) [right of=3] {$4$};
\draw (0) -- (3); 
\draw (0) -- (4); 
\draw (1) -- (3); 
\draw (1) -- (4);
\draw (2) -- (3);
\draw (2) -- (4);
\draw (3) -- (4); 
\end{tikzpicture} }
        \caption{Exp 18 Graph}
    \end{subfigure}
    \begin{subfigure}[t]{0.24\textwidth}
        \raisebox{-\height}{\begin{tikzpicture}[node distance={15mm}, thick, main/.style = {draw, circle}] 
\node[main] (0) {$0$}; 
\node[main] (1) [above right of=0] {$1$}; 
\node[main] (2) [right of=1] {$2$}; 
\node[main] (3) [below right of=0] {$3$}; 
\node[main] (4) [right of=3] {$4$};
\draw (0) -- (4); 
\draw (1) -- (4); 
\draw (2) -- (4); 
\end{tikzpicture}  }
        \caption{Exp 3 Graph}
    \end{subfigure}
    \hfill
    \begin{subfigure}[t]{0.24\textwidth}
        \raisebox{-\height}{\begin{tikzpicture}[node distance={15mm}, thick, main/.style = {draw, circle}] 
\node[main, fill=lightgray] (1) {$1$}; 
\node[main, fill=lightgray] (0) [above right of=1] {$0$}; 
\node[main, fill=lightgray] (2) [right of=0] {$2$}; 
\node[main, fill=lightgray] (3) [below right of=1] {$3$}; 
\node[main, fill=lightgray] (4) [right of=3] {$4$};
\draw (0) -- (2); 
\draw (0) -- (3); 
\draw (0) -- (4); 
\draw (2) -- (3);
\draw (2) -- (4);
\draw (3) -- (4); 
\end{tikzpicture} }
        \caption{Exp 23 Graph}
    \end{subfigure}
    \hfill
    \begin{subfigure}[t]{0.24\textwidth}
        \raisebox{-\height}{\begin{tikzpicture}[node distance={15mm}, thick, main/.style = {draw, circle}] 
\node[main] (2) {$2$}; 
\node[main] (0) [above right of=2] {$0$};  
\node[main] (3) [below right of=2] {$3$}; 
\node[main] (1) [right of=3] {$1$};
\node[main] (4) [right of=0] {$4$};
\draw (0) -- (3); 
\draw (0) -- (4); 
\draw (1) -- (3); 
\draw (1) -- (4); 
\end{tikzpicture}}
        \caption{Exp 13 Graph}
    \end{subfigure}
    \hfill
    \begin{subfigure}[t]{0.24\textwidth}
        \raisebox{-\height}{\begin{tikzpicture}[node distance={15mm}, thick, main/.style = {draw, circle}] 
\node[main, fill=lightgray] (0) {$0$}; 
\node[main, fill=lightgray] (1) [above right of=0] {$1$}; 
\node[main, fill=lightgray] (2) [right of=1] {$2$}; 
\node[main, fill=lightgray] (3) [below right of=0] {$3$}; 
\node[main, fill=lightgray] (4) [right of=3] {$4$};
\draw (0) -- (1); 
\draw (0) -- (2); 
\draw (0) -- (3); 
\draw (0) -- (4); 
\draw (1) -- (2); 
\draw (1) -- (3); 
\draw (1) -- (4); 
\draw (2) -- (3);
\draw (2) -- (4);
\draw (3) -- (4); 
\end{tikzpicture} }
        \caption{Exp 33 Graph}
    \end{subfigure}
    \caption{Graphs of the chosen experiments. The light gray color is only chosen to better distinguish different graphs.}
    \label{fig:graphs}
\end{figure}
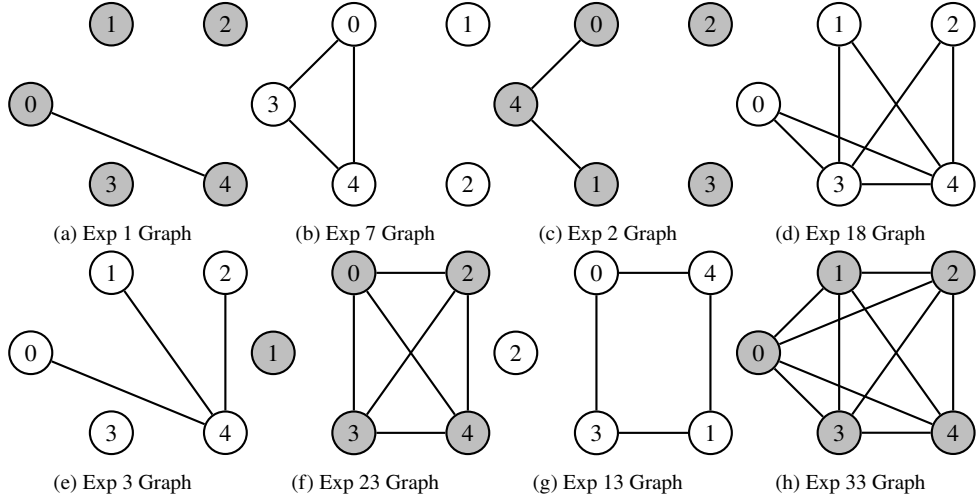
\begin{figure}
    \centering
    \begin{subfigure}[t]{0.49\textwidth}
        \raisebox{-\height}{\includegraphics[width=\textwidth]{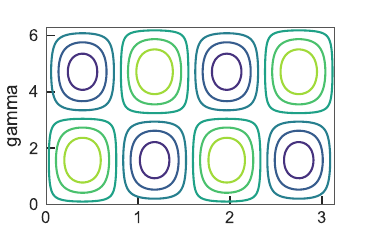}}
        \caption{Exp 1 Energy Landscape}
        \label{fig:exp1}
    \end{subfigure}
    \hfill
    \begin{subfigure}[t]{0.49\textwidth}
        \raisebox{-\height}{\includegraphics[width=\textwidth]{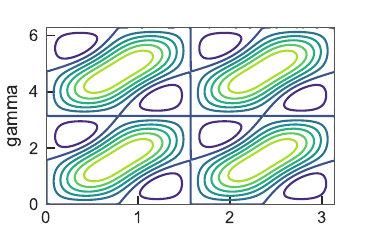}}
        \caption{Exp 7 Energy Landscape}
        \label{fig:exp7}
    \end{subfigure}
    \begin{subfigure}[t]{0.49\textwidth}
        \raisebox{-\height}{\includegraphics[width=\textwidth]{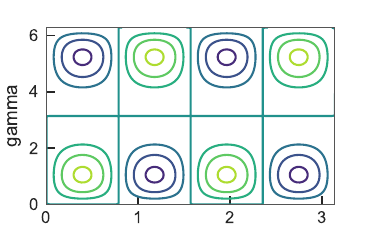}}
        \caption{Exp 2 Energy Landscape}
        \label{fig:exp2}
    \end{subfigure}
    \hfill
    \begin{subfigure}[t]{0.49\textwidth}
        \raisebox{-\height}{\includegraphics[width=\textwidth]{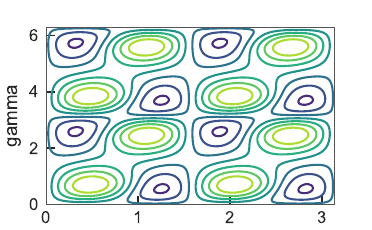}}
        \caption{Exp 18 Energy Landscape}
        \label{fig:exp18}
    \end{subfigure}
    \begin{subfigure}[t]{0.49\textwidth}
        \raisebox{-\height}{\includegraphics[width=\textwidth]{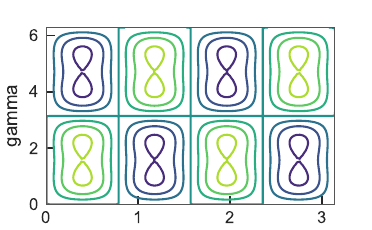}}
        \caption{Exp 3 Energy Landscape}
        \label{fig:exp3}
    \end{subfigure}
    \hfill
    \begin{subfigure}[t]{0.49\textwidth}
        \raisebox{-\height}{\includegraphics[width=\textwidth]{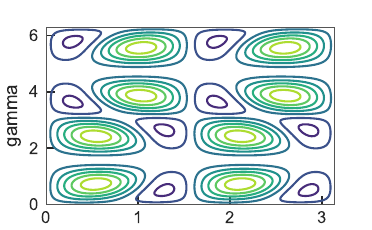}}
        \caption{Exp 23 Energy Landscape}
        \label{fig:exp23}
    \end{subfigure}
    \begin{subfigure}[t]{0.49\textwidth}
        \raisebox{-\height}{\includegraphics[width=\textwidth]{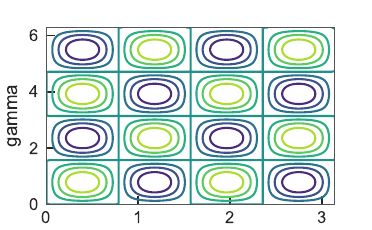}}
        \caption{Exp 13 Energy Landscape}
        \label{fig:exp13}
    \end{subfigure}
    \hfill
    \begin{subfigure}[t]{0.49\textwidth}
        \raisebox{-\height}{\includegraphics[width=\textwidth]{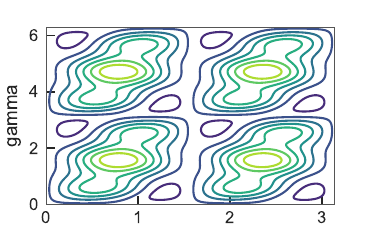}}
        \caption{Exp 33 Energy Landscape}
        \label{fig:exp33}
    \end{subfigure}
    \caption{Energy landscapes of the chosen experiments. Within the energy landscapes, lighter/yellow colors indicate high values and darker/blue colors indicate lower values. The data of these plots is available at GitHub \url{https://github.com/danielmpla/sparseq/commit/6ddc7beefe111744bfa6c6219ee28e380344b618}}
    \label{fig:landscapes}
\end{figure}
Additional observations highlight specific changes and their impacts. Exp 2 (\cref{fig:exp2}) adds an edge on node 4 compared to Exp 1, creating more interaction and a slightly more complex energy landscape. Exp 3 (\cref{fig:exp3}) adds another edge to node 4, further increasing the complexity. Exp 13 (\cref{fig:exp13}) introduces a four-node cycle, which significantly changes the energy landscape by adding more interconnected pathways, leading to a more complex and interactive pattern.
In \cref{fig:landscapes}, more complex graphs such as Exp 18 (\cref{fig:exp18}), Exp 23 (\cref{fig:exp23}), and Exp 33 (\cref{fig:exp33}) feature a higher number of three-node subgraphs due to increased connectivity and a denser network of edges, significantly contributing to the complexity and interconnectivity of these graphs. The energy landscapes of these graphs display numerous interacting energy minima and saddle points. For example, Exp 18, with seven edges, shows a highly interconnected pattern with many interacting energy minima, suggesting multiple interaction pathways. Exp 23, with even more edges, exhibits a very complex pattern with numerous energy minima and saddle points, indicating a dense network and an intricate energy landscape. The fully connected graph in Exp 33 exhibits the most complex energy landscape, characterized by tightly coupled and densely populated contours, indicating maximal interaction pathways and a highly intricate energy distribution.


Compared to the work by St{\k e}ch{\l}y et al. \cite{stechly_connecting_2023}, our roughness metrics given in \cref{tab:experiments}, however, are smaller, indicating that the sparsity has less effect on the roughness than the size of the Hamiltonian coefficients.

\section{Implications on QAOA}
\label{sec:implications}
In this section, we analyze how sparsity affects the optimization performance for QAOA on a simulator.
We implemented QAOA to solve the Max-Cut problem for specific graphs. This setup involves initializing our custom QAOA class with parameters such as the quantum operator and offset. These parameters were generated with IBMs DOCPLEX by converting our quadratic program using \cref{eq:max-cut} into an ising Hamiltonian. The algorithm runs on the IBM Q experience platform, using the QASM simulator with 2048 shots to perform the quantum circuit executions. Our QAOA class includes methods to calculate costs, compute expectation values from measurement outcomes, and create parameterized quantum circuits based on the QAOA ansatz. We used a single layer for QAOA to ensure that our results are comparable with the analytic cost function.
In our experimental setup, we selected the experiments from \cref{tab:experiments}. Each instance was processed using the QAOA algorithm, where the optimization process was driven by the Simultaneous Perturbation Stochastic Approximation (SPSA) optimizer with a maximum of 1000 iterations. The expectation values were computed on the basis of measurement counts from simulated quantum executions.
In general, we found that our simulations demonstrated performance on par with the analytical cost function, indicating a high degree of accuracy and reliability. Furthermore, the SPSA optimizer effectively found the minimum without encountering significant issues, demonstrating its robustness and efficiency in this context. Consequently, the performance of the QAOA appears to be uncoupled from the sparsity of the problem Hamiltonian, suggesting that the algorithm's efficacy is not significantly influenced by the sparsity characteristics of the problem Hamiltonian.

\section{Conclusion and Outlook}
\label{sec:conclusion}
This study delves into the intricate relationship between the sparsity of the augmented interaction matrices and the performance of QAOA in solving combinatorial optimization problems, specifically focusing on the Max-Cut problem. Through both analytical and numerical analyses, we have shown that the sparsity of the augmented interaction matrix significantly impacts the energy landscape and but not so much the optimization efficiency of QAOA. 
Our findings suggest that, while denser matrices introduce greater complexity into the energy landscape, leading to numerous local minima and saddle points, the overall performance of QAOA does not strictly degrade with decreasing sparsity. This is evidenced by similar optimization results across various sparsity levels in our experiments.
The analytical derivation of cost functions for different sparsity levels provided deeper insights into how the connectivity and structure of the problem Hamiltonian influence the roughness metrics, such as Total Variation and Fourier Density. Comparing our values of these metrics with the values in \cite{stechly_connecting_2023} indicates that the size of the coefficients has a more pronounced effect on the roughness of the energy landscape than the sparsity itself.
Looking forward, future research should explore the application of QAOA to larger and more complex problem instances, potentially involving real-world datasets. Investigating the interaction between the sparsity of the augmented interaction matrix and other algorithmic parameters, such as circuit depth and noise levels, could yield further enhancements in QAOA performance. 
In conclusion, while the sparsity of the enhanced interaction matrix alone presents a significant factor in the structure of the energy landscape, the adaptability and robustness of QAOA, combined with effective optimization strategies, can mitigate many of the associated challenges.

\printbibliography

\section*{Acknowledgements}
Funded by the European Union under Horizon Europe Programme - Grant Agreement 101080086 — NeQST.\\
Views and opinions expressed are however those of the author(s) only and do not necessarily reflect those of the European Union or European Climate, Infrastructure and Environment Executive Agency (CINEA). Neither the European Union nor the granting authority can be held responsible for them.\\
We acknowledge the use of OpenAI's ChatGPT (GPT-4) for assisting with the data analysis and generating initial drafts of several sections of this manuscript. ChatGPT helped streamline our research process by providing valuable insights and expediting the writing phase.

\end{document}